\documentclass[pra,aps,reprint,a4paper,superscriptaddress,notitlepage,preprintnumbers,longbibliography]{revtex4-1}
\usepackage{graphicx}
\usepackage{dcolumn}
\usepackage{bm}
\usepackage{amsmath}
\usepackage{amssymb}
\usepackage{booktabs}
\usepackage{color,xcolor}
\usepackage{epsfig}
\usepackage[pdfstartview=FitH,linkbordercolor = white,colorlinks = true,linkcolor=blue,citecolor=red,urlcolor=green,urlcolor=blue]{hyperref}
\newcommand{\redc}[1]{\textcolor{black}{#1}}

\newcommand{\ket}[1]{|#1\rangle}
\newcommand{\braket}[2]{\langle #1|#2\rangle}
\newcommand{\Ignore}[1]{}
\newcommand{\im}{{\rm Im}}
\newcommand{\re}{{\rm Re}}
\newcommand{\msc}[1]{\text{\textsc{#1}}}
\newcommand{\NMP}{N}
\newcommand{\Pm}{\phantom{-}}
\def\fisunical{Dip. di Fisica, Universit\`{a} della Calabria, Via P. Bucci, Cubo 30C, I-87036 Rende (CS), Italy}
\def\infnlnf{INFN, sezione LNF, Gruppo collegato di Cosenza, Cubo 31C, I-87036 Rende (CS), Italy}
\def\UAM{Dept. de Qu\'{\i}mica, Univ. Aut\'{o}noma de Madrid, C/ Fco. Tom\'{a}s y Valiente 7, E-28049 Madrid, Spain}
\begin{document}

\title{Plasmon properties and hybridization effects in Silicene}
\author{\firstname{C.}~\surname{Vacacela Gomez}}
\affiliation{\fisunical}
\affiliation{\infnlnf}

\author{\firstname{M.}~\surname{Pisarra}}
\affiliation{\UAM}

\author{\firstname{M.}~\surname{Gravina}}
\affiliation{\fisunical}
\affiliation{\infnlnf}

\author{\firstname{P.}~\surname{Riccardi}}
\affiliation{\fisunical}
\affiliation{\infnlnf}

\Ignore{
\author{\firstname{J. M.}~\surname{Pitarke}}
\email{jm.pitarke@nanogune.eu}
\affiliation{CIC nanoGUNE, Tolosa Hiribidea 76, E-20018 Donostia--San Sebastian, Basque Country, Spain}
\affiliation{Materia Kondentsatuaren Fisika Saila, DIPC, and Centro Fisica Materiales CSIC-UPV/EHU, 644 Posta Kutxatila, E-48080 Bilbo, Basque Country, Spain}
}

\author{\firstname{A.}~\surname{Sindona}}
\email{antonello.sindona@fis.unical.it}
\affiliation{\fisunical}
\affiliation{\infnlnf}

\begin{abstract}
The plasmonic character of monolayer silicene is investigated by time-dependent density functional theory in the random phase approximation. Both the intrinsic (undoped) and several extrinsic (carrier doped or gated) conditions are explored by simulating injection of a probe particle (i.e., an electron or a photon) of energy below 20~eV and in-plane momentum smaller than 1.1~\AA$^{−1}$. 
The energy-loss function of the system is analyzed, with particular reference to its induced charge-density fluctuations, i.e., plasmon resonances and corresponding dispersions, occurring in the investigated energy-momentum region. 
At energies larger than $1.5$~eV, two intrinsic interband modes are detected and characterized. The first one is a hybridized $\pi$-like plasmon, which is assisted by competing one-electron processes involving sp$^2$ and sp$^3$ states, and depends on the slightest changes in specific geometric parameters, such as nearest-neighbor atomic distance and buckling constant. 
The second one is a more conventional $\pi$-$\sigma$ plasmon, which is more intense than the $\pi$-like plasmon and more affected by one-electron processes involving the $\sigma$ bands with respect to the analogous collective oscillation in monolayer graphene. 
At energies below $1$~eV, two extrinsic intraband modes are predicted to occur, which are generated by distinct types of Dirac electrons (associated with different Fermi velocities at the so-called Dirac points). 
The most intense of them is a two-dimensional plasmon, having an energy-momentum dispersion that resembles that of a two-dimensional electron gas. 
The other is an acoustic plasmon that occurs for specific momentum directions and competes with the two-dimensional plasmon at mid-infrared energies. 
The strong anisotropic character of this mode cannot be explained in terms of the widely used Dirac-cone approximation. 
As in mono-, bi-, and few-layer graphene, the extrinsic oscillations of silicene are highly sensitive to the concentration of injected or ejected charge carriers.
More importantly, the two-dimensional and acoustic plasmons appear to be a signature of the honeycomb lattice, independently of the chemistry of the group-IV elements and the details of the unit-cell geometry.
\end{abstract}


\maketitle

\section{Introduction}
Atomically thin materials, organized in hexagonal honeycomb like geometry, have been through intense scrutiny due to their exceptional electronic properties, the most intriguing of which is, perhaps, the strong coupling between their quantized charge-density fluctuations, i.e., plasmon modes, and light or other charged particles~\cite{lin2008operation,ju2011graphene,bao2012graphene,sensale2012broadband,garcia2014graphene}.
Control and manipulation of the plasmon properties of these two-dimensional systems\redc{, including graphene and beyond,} are expected to guide the design of next-generation nanophotonic and nanoelectronic devices, with the enhanced capability to operate from terahertz~(THz) to infrared~(IR) frequencies.
\redc{Indeed, plasmon-like modes appear as the ``true'' low-energy excitations of low-dimensional system~\cite{haldane_1,haldane_2}, while charged and spinful modes are realized as coherent states, with their own peculiar dynamics~\cite{giula_1,giula_2}, both in normal, and in superconducting phases~\cite{moscio,giuso_1,giuso_2,1367-2630-12-2-025022}.}
Fundamental research and technological applications have been historically focused on monolayer graphene~(MG)~\cite{GeimNovGraphene2007,neto2009electronic}, whose extrinsic plasmons exhibit much stronger confinement, larger tunability and lower losses compared to gold and silver nanoparticles~\cite{christensen2011graphene}.
Recent studies have thoroughly characterized the plasmon dispersions in MG~\cite{grigorenko2012graphene,pisarra2014acoustic,liou2015pi,woessner2015highly,0022-3727-48-46-465104,sindona2016plasmon} and some other graphene-related materials, such as bilayer graphene~(BLG)~\cite{FeiBLG2015,pisarra2016dielectric} and graphene nanoribbons~(GNRs) arranged in regular planar arrays,\redc{ which, unlike MG, offer geometrically controllable band gaps}~\cite{fei2015edge,vacacela2016gnr,vacacelaBJNANO2017}.
\Ignore{A band gap can be hardly opened in the band structure of MG to a sizable value.}
Further advances are expected from the analysis of nanocarbon-metal interfaces together with associated application developments in biological sensing, optical signal processing and quantum information technology~\cite{AnkerNat2008,bao2012graphene,PhysRevA.92.053812,TsargorodskaNL2016,acssensors5b00280,Genslein2016}.

Nonetheless, the incompatibility of carbon-based materials with current silicon-based electronics makes the former unsuitable for immediate use, for which reason, other candidates among the group-IV elements, e.g., Silicon and Germanium in honeycomb lattice, have been explored~\cite{houssa2011electronic,vogt2012silicene,drummond2012electrically,kara2012review,o2012stable,cai2013stability,zhao2016rise,kaloni2016current} \redc{ that, similarly to MG, have, in their free-standing forms, linear dispersing valence~($\pi$) and conduction~($\pi^{*}$)~bands, crossing at the inequivalent  K (and K$^{\prime}$) points~\cite{cahangirov2009two,liu2011quantum}, where charge-carries behave as massless Dirac fermions.}

In particular, silicene has been recognized to have four principal advantages with respect to MG:
(i) compatibility with current electronics;
(ii) \redc{large spin-orbit  induced band gap of $1$-$10$~meV~\cite{liu2011quantum,PhysRevB.84.195430}, with respect to the $10^{-3}$~meV value predicted for MG~\cite{PhysRevB.75.041401}, which has been confirmed by an experimental realization of the quantum spin Hall effect~\cite{PhysRevLett.112.106802};}
(iii) tunable band gap in the presence of a perpendicular electric field~\cite{ni2011tunable,PhysRevB.89.195410,PhysRevB.89.201411,PhysRevB.90.035142,drummond2012electrically}, \redc{with band gap values of several meV that increase with increasing the field strength}, or by hydrogenation~\cite{Osborn2011101,Lu2009153,Shyam2014} and fluorination~\cite{Ding2012}; and
(iv) synthesis on various substrates with different electronic properties~\cite{zhao2016rise}.
This and other beyond-graphene systems are expected to be good competitors with  graphene-related  materials\redc{, sharing with the latter many of the superior features of MG, plus enhanced controllable electronic properties for specific applications.}
A further support in this direction has been the fabrication of a \Ignore{first} silicene-based field-effect-transistor~\cite{tao2015silicene}.

From the structural point of view, atomic resolved scanning tunneling microscopy~(STM) experiments, combined with density-functional calculations~\cite{vogt2012silicene,lin2012structure,feng2012evidence,Shyam2014},  have demonstrated that silicene is characterized by a buckled honeycomb lattice where the sp$^{2}$ hybridized orbitals get slightly dehybridized into sp$^{3}$~like orbitals causing a weakening of the $\pi$ bonds.
This mixed sp$^{2}$-sp$^{3}$ hybridization yields a nearest-neighbor Si-Si distance of $2.2$-$2.3$~{\AA} and a low buckling of $0.45$-$0.55$ {\AA}.

On the practical side, silicene has been successfully grown on transition metals, such as Ag(001),  Ag(110), Ag(111)~\cite{kara2012review,zhao2016rise,kaloni2016current},  Ir(111)~\cite{meng2013buckled}, metal dichalcogenides, like MoS$_2$~\cite{gao2014tunable,Qian1256815}, and ceramic materials, e.g., ZrB$_2$(0001)~\cite{fleurence2012experimental}.
Combined experimental and theoretical studies have demonstrated that the Dirac cone disappears in silicene grown on Ag substrates, due to strong hybridization between Si and Ag orbitals~\cite{lin2013substrate,guo2013absence,cahangirov2013electronic,PhysRevB.89.201416}.
Further theoretical scrutiny has also predicted silicene  to be energetically stable (with unperturbed Dirac cones) on h-BN, H-passivated C-SiC(0001), Si-SiC(0001), Cl-passivated Si(111), CaF$_2$, AlAs(111), AlP(111), GaAs(111), GaP(111), ZnSe(111), and ZnS(111)~\cite{kaloni2016current,guo2013absence,kaloni2013quasi,liu2013silicene,kokott2014nonmetallic,noguchi2015direct}. Semiconducting substrates with large band gaps like h-BN turn out to be excellent candidates to adsorb pristine silicene,
simply because no substrate states appear close to the Fermi level and,
accordingly, the interaction silicene-substrate cannot perturb the Dirac-cone structure.

Although the structural properties of pristine silicene are well established, a comprehensive approach and description of its plasmon properties at low and high energies, from the IR to ultraviolet~(UV) range, has not been presented so far.
A few theoretical studies~\cite{mohan2013first,das2015optical,matthes2013universal,matthes2014optical}, based on the analysis of the electron energy-loss function and absorption spectrum, have reported the existence of two interband plasmons in intrinsic silicene, occurring at energies larger than $1.5$~eV for vanishingly small momentum transfers.
These are counterparts to the well-known $\pi$ and $\pi$-$\sigma$ plasmons found in MG, BLG and graphite~\cite{eberlein2008plasmon}.
Otherwise, MG, silicene and germanene have been proved to have the same IR frequency absorbance~\cite{bechstedt2012infrared}, which is a universal feature of the two-dimensional  honeycomb lattice and does not depend on the chemistry of the group-IV element, buckling parameter or hybridization state.
\redc{Furthermore, some noteworthy approaches have investigated the effect of an external, perpendicular electric field~\cite{PhysRevB.89.195410,PhysRevB.89.201411}, plus an additional exchange field~\cite{PhysRevB.90.035142}, probing the spin-orbit band-structure of silicene, below some tenths of eV around the K and K$^{\prime}$ points.
In these studies, the use of an effective two-spinor Hamiltonian, with an emergent Dirac-cone structure perturbed by spin-orbit~\cite{PhysRevLett.95.226801} and external/field terms, has revealed presence and interplay of \Ignore{two undamped} extrinsic plasmons, at meV energies and far-IR wave lengths, displaying  features  of  different regimes,  including those of a topological insulator and a valley-spin polarized metal.
In the same energy-momentum region, of key-interest for nanodevice applications, it has been demonstrated that the electron-phonon interaction plays a critical role~\cite{PhysRevB.88.121403,WangJAP2015}, leading the in-plane phonon modes to hybridize with the extrinsic plasmons.}

\redc{Here, we present a full {\it ab initio} approach, based on time-dependent~(TD) density functional theory~(DFT) within the random phase approximation~(RPA), to scrutinize the plasmon properties of pristine and doped silicene in absence of external fields and spin-orbit effects, which makes an excellent approximation for  applications in the mid-IR to UV range, with the plasmon-phonon coupling reduced to a minor extent.}
\redc{First, we focus on the intrinsic IR and UV plasmons, whose long-wavelength behavior is in agreement with the absorbance predictions of Refs.~\onlinecite{matthes2013universal} and~\onlinecite{matthes2014optical}.
In particular, we describe the role played by sp$^2$ and sp$^3$ hybridization in the Landau damping mechanism, which affects the silicene counterpart of the MG $\pi$ plasmon.
Next, we discuss the mid-IR to near-IR extrinsic plasmons, highlighting the occurrence of two distinct tunable modes, which may be seen as a universal feature of the gapless band-structure of group IV elements in honeycomb lattice.}

Our arguments are organized as follows.
In section~II, we briefly outline the key concepts of our TDDFT+RPA scheme, suitable for two-dimensional  periodic systems.
In section~III, we \Ignore{present and} analyze the  energy-loss  spectrum of silicene at high energies~(larger than $\sim 1.5$~eV), with reference to the coupling of the intrinsic plasmons and single-particle (SP) processes, involving sp$^2$ and sp$^3$ hybridized states.
In section~IV, we investigate the low-energy end of the  energy-loss  spectrum~(below $\sim 1$~eV) in the presence of carrier doping, and show the interrelation between two extrinsic intraband modes: one of them is a two-dimensional plasmon, already observed in a variety of graphene-based materials; the other is an acoustic plasmon, direct consequence of the anisotropic band structure of the system around the Dirac points, i.e., the failure of the Dirac-cone approximation.
In section V, we draw our conclusions.

\section{Methods}
In this section, we first provide the input parameters of our ground-state density functional calculations of monolayer silicene~(Sec.~\ref{DFTCalcs}).
Then, we highlight the main features of our TDDFT approach, yielding the complex permittivity and loss function of the system under both intrinsic and extrinsic conditions~(Sec.~\ref{3Dvs2Dapproaches}).
Hartree atomic units are used throughout, unless otherwise stated.
All calculations were performed  on a high performance computing cluster~\cite{GALILEO-SCAI}, using up to 512 CPUs on 32 nodes with 3840 Gb memory.

\subsection{DFT calculation\label{DFTCalcs}}
\begin{figure}[t]
\centering{
\includegraphics[width=0.49\textwidth]{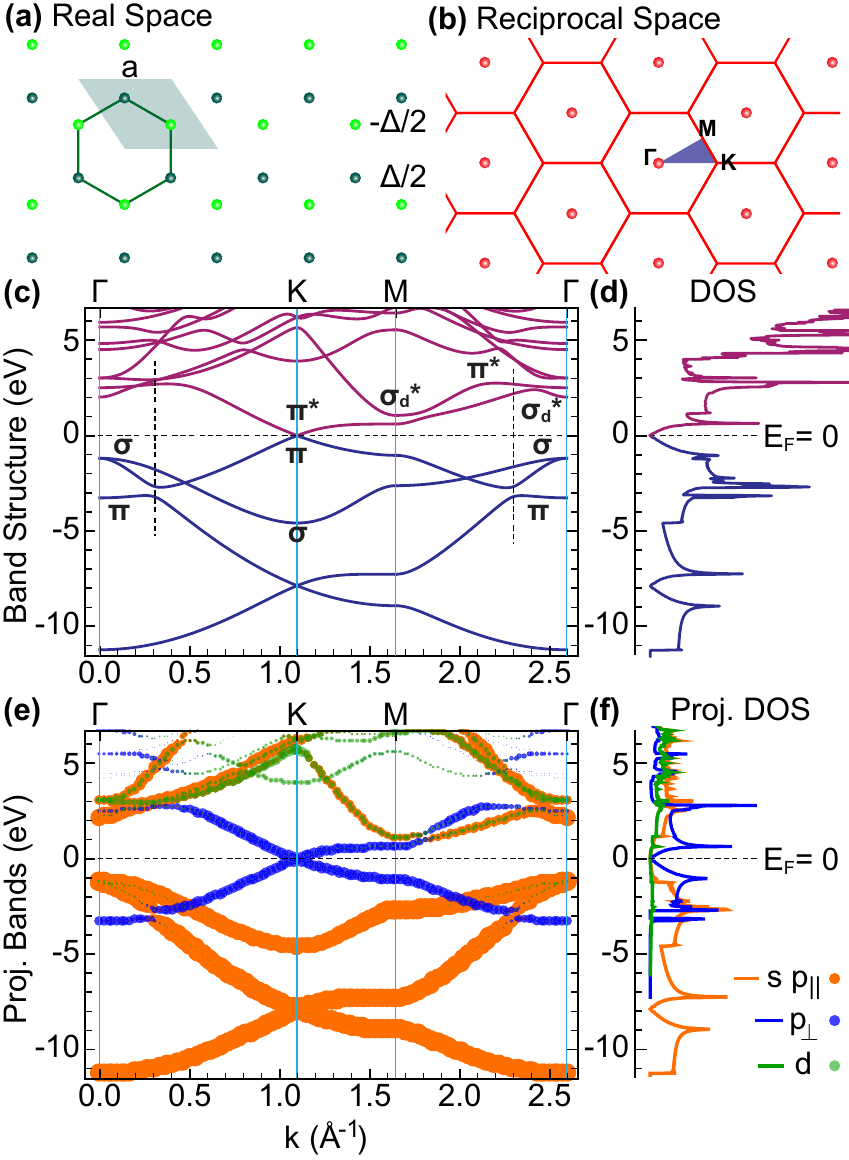}
}
\vskip -14pt
\caption{Geometric and ground state properties of silicene: (a) direct lattice, unit cell~(green shaded rhomboid), lattice constant $a=3.82$~{\AA}, and buckling parameter $\Delta=0.45$~{\AA}; (b) reciprocal lattice, first BZ, irreducible BZ~(blue shaded triangle), and high-symmetry points ${\Gamma}$, $K$, $M$; (c) band structure along ${\Gamma}KM{\Gamma}$, with the Fermi level set to zero energy;
(d) DOS profile corresponding to the energy width of panel~(c);
(e) orbital projected band structure along ${\Gamma}KM{\Gamma}$  onto $s\,p_{\parallel}$, $p_{\perp}$ and $d$ states of the unit-cell atoms (with $\parallel$ denoting the in-plane bons and $\perp$ the out-of-plane bonds). The radius of each point is proportional to the weight of the corresponding atomic contribution; (f) orbital projected DOS corresponding to the orbital projected bands of panel~(e).
\label{bandSi}
}
\end{figure}

The equilibrium electronic properties of silicene were determined by DFT in the local density approximation~(LDA)~\cite{perdew1981self}, with the Kohn-Sham~(KS) electron wave functions expanded in the plane-wave~(PW) basis~\cite{gonze2009abinit}.
The latter is represented by the space functions
\begin{equation}
{\rm PW}_{{\bf k}+{\bf G}}({\bf r})=\Omega_0^{-1/2}e^{ i  ({\bf k}+{\bf G})\cdot {\bf r}}, \label{PW}
\end{equation}
where ${\bf k}$ is a wave vector in the first Brillouin zone~(BZ),  ${\bf G}$ a reciprocal-lattice vector, and  ${\Omega_0}$ the unit-cell volume associated to the real-space lattice~[Fig.~\ref{bandSi}(a) and \ref{bandSi}(b)].
The number of PWs was limited by the energy cut off $|{\bf k}+{\bf G}|/2 \leq 25$~Hartrees.
Norm conserving pseudopotentials of the Troullier-Martins type were adopted to eliminate the core electrons~\cite{troullier1991efficient}.
The three-dimensional periodicity inherent in our PW-DFT approach was generated by replicating  the silicene planes with a minimum separation $L$ of $20$~{\AA}.
\Ignore{This value ensures negligible overlap~(but not negligible interaction) of charge densities located at the different replicas.}

Geometry optimization and ground state calculations were carried out on the irreducible part of the first BZ [$\Gamma{KM}$ triangle in Fig.~\ref{bandSi}(b)], using a $\Gamma$-centered and unshifted Monkhorst-Pack~(MP) grid of ${\NMP}=180{\times}180{\times}1$ ${\bf k}$ points~\cite{monkhorst1976special}.
The optimized lattice constant $a$ and buckling parameter $\Delta$ were found to be $a=3.82$~{\AA} and $\Delta=0.45$~{\AA}, respectively~[Fig.~\ref{bandSi}(a)].
Slightly larger lattice-constant values available from the literature~\cite{liu2011quantum,liu2013silicene,BalendhranSmall2014,Shyam2014} were also tested, while keeping the buckling parameter fixed to its optimized value~(see Appendix~\ref{appI}).

As for the LDA electronic structure, the KS energies $\varepsilon_{\nu {\bf k}}$  and wave functions
\begin{equation}
\braket{{\bf r}} {\nu {\bf k}} =
{\NMP}^{-1/2}\sum_{\bf G} c_{\nu {\bf k}+{\bf G}} {\rm PW}_{{\bf k}+{\bf G}}({\bf r}),
\label{wfnk}
\end{equation}
were computed for $\nu \leq 50$ bands, with the ${\bf G}$ sum being limited by the energy cut off condition set forth above to include $\sim 5000$ coefficients $c_{\nu {\bf k}+{\bf G}}$ per wave function.

A simple visual perspective of the band energies~[Fig.~\ref{bandSi}(c)] shows that silicene presents a Dirac-cone structure at the $K$ points, within $\sim 0.8$~eV around the Fermi energy $E_F$.
Looking at the density of states~(DOS) as function of the band energies~[Fig.~\ref{bandSi}(d)], we notice that the two peaks closest to $E_F$ correspond to $\pi$ and $\pi^*$ flat band dispersions at the $M$ point~[Fig.~\ref{bandSi}(c)].
Other two peculiar DOS-peak structures, appearing at $\sim3$-$4$~eV above and below $E_F$, are direct consequence of the mixed sp$^{2}$-sp$^{3}$ hybridization and buckled conformation of silicene.\Ignore{, which is manifested in avoided crossing points between the (bonding) $\sigma$ and $\pi$ bands and the (anti-bonding)  $\sigma^*$ and $\pi^*$ bands~\cite{Note1}.}
A more-detailed insight into these hybridization mechanisms is offered by the orbital projected band structure and DOS of Fig.~\ref{bandSi}(e) and~\ref{bandSi}(f),  respectively, where we have separated the contribution of $s$ and $p_{\parallel}$ states, forming $sp^{2}$~like bonds, from $p_{\perp}$ states, involved in $\pi$ bonds, and  $d$ states of the unit-cell Si atoms.
The valence bands have well-defined $\sigma$ and $\pi$ characters, with sharp avoided-crossing features in correspondence of the DOS peaks at $2.5$-$3.5$~eV below $E_F$.
The conduction states above the Dirac point are also of the $\pi$ form, yielding an antibonding $\pi^*$ band responsible for the DOS peak at $2.7$~eV above $E_F$~\cite{cahangirov2009two}.
Another conduction band denoted $\sigma_d^*$ lies close to the $\pi^*$ band and produces a DOS peak at $3$~eV above $E_F$. It has a dominant $\sigma$-character contaminated by $d$ states.
Other conduction bands are strongly influenced by $d$, $f$, and higher principal quantum numbers, as it can be deduced by comparing the full DOS and its projected $s\,p_{\parallel}$, $p_{\perp}$ and $d$ components at energies larger than $\sim 3$ eV above $E_F$.
\Ignore{Other peculiar DOS peaks, in the range of $\pm4$~eV from $E_F$, originate from the mixed sp$^2$-sp$^3$ hybridization of buckled silicene, and hence they are absent in MG.}
\Ignore{In particular, the small \Ignore{or vanishing} difference in energy between the $\sigma^{*}$ and $\pi^{*}$ bands at $\sim 2.7$~eV above $E_F$ is associated to several quasiflat band-dispersion points, which induce a larger DOS peak than... .}
\Ignore{Though the $\pi$ and $\sigma$ character of the excited energy levels above $E_F$ is not guaranteed away from the $K$ and $M$ points~\cite{PhysRevB.89.165430}, we shall continue to refer to them as being part of the $\pi^*$ and $\sigma^*$ bands, for notational simplicity.}

These characteristics~(not found in  MG\Ignore{, where a comprehensive group classification of the band structure is available}~\cite{PhysRevB.89.165430}) play an important role in the plasmonic properties of silicene at probing energies larger than $1.5$~eV, as we will see in Sec.~\ref{IntrinsicSil}.
\begin{figure}[t]
\centering{
\includegraphics[width=0.45\textwidth]{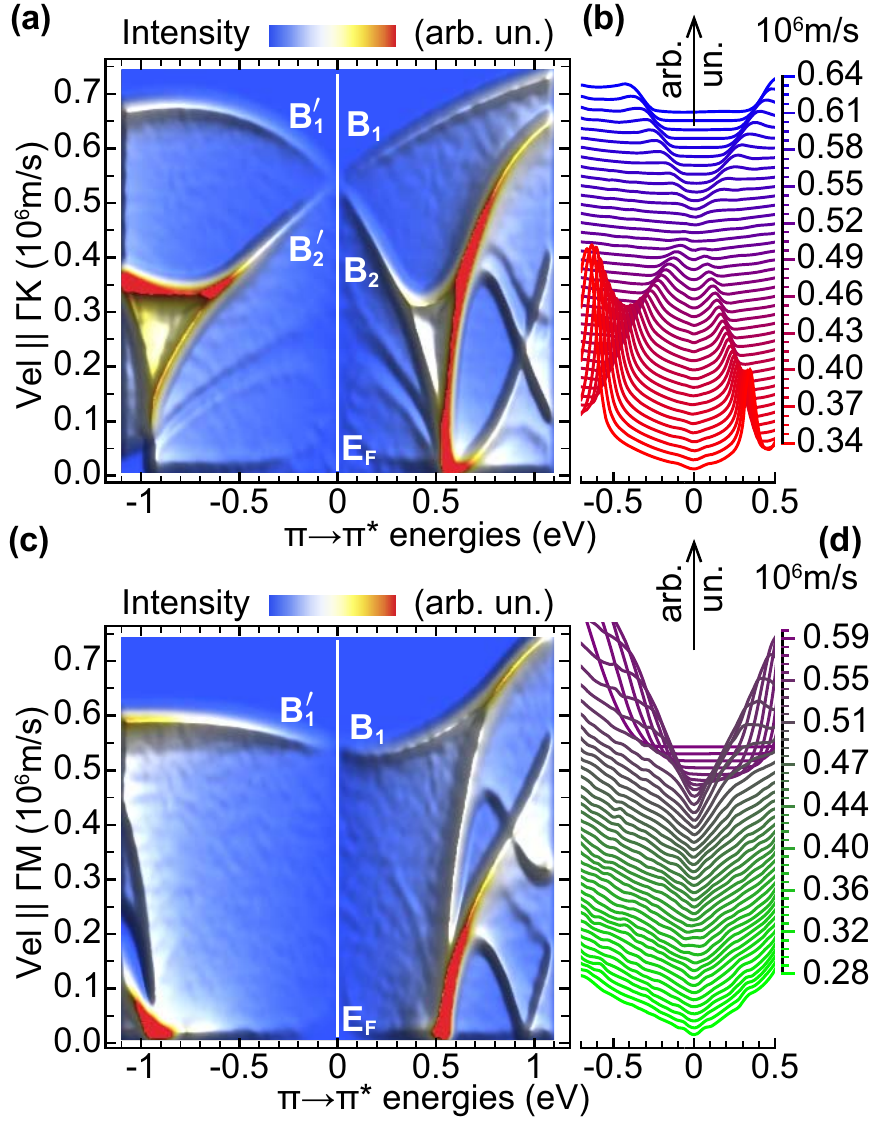}
}
\vskip -14pt
\caption{Partial DOS vs the one-electron energy of the $\pi$ and $\pi^*$ bands and group velocity along ${\Gamma}K$ [(a),(b)]  and ${\Gamma}M$ [(c),(d)].
The intensity scale in the density plots of (a) and (c) is cut at $12\%$ of the peak maximum, occurring for small group velocities, to emphasize the behavior of the Dirac electrons, whose intensity peaks are denoted B$_1$, B$_1^{\prime}$, B$_2$, B$_2^{\prime}$.
(b) and (d) provide a complementary view of (a) and (c), respectively, with the partial DOS curves being reported for fixed group velocity values and shifted vertically for clarity.
\label{FVeloc}
}
\end{figure}

As for the Dirac point  features of the $\pi$-$\pi^*$ electrons, an average Fermi-velocity value  $v_F = 0.54{\times} 10^{-6}$~m/s can be extracted from the band data, which is roughly $65\%$ of that derived from local density calculations in MG\redc{, and reflects a relatively reduced mobility of the massless Dirac fermions of Silicene\Ignore{with energies close to the Fermi level}.}
A more detailed analysis is made possible by inspection of the partial DOS as function of the $\pi$-$\pi^*$ energies and group velocities parallel to $\Gamma$K~[Figs.~\ref{FVeloc}(a) and \ref{FVeloc}(b)] and $\Gamma$M~[Fig.~\ref{FVeloc}(c) and \ref{FVeloc}(d)].
Along $\Gamma$K, the DOS is peaked at two distinct Fermi-velocity values, corresponding to the B$_1^{\prime}$, B$_2^{\prime}$ branches below $E_F$ and the B$_1$, B$_2$ branches above $E_F$ in Fig.~\ref{FVeloc}(a).
On the other hand, along $\Gamma$M, the DOS is peaked around one single Fermi-velocity value, associated to  the B$_1^{\prime}$ branch below $E_F$ and B$_1$ branch above $E_F$ in Fig.~\ref{FVeloc}(c).

Like in MG~\cite{pisarra2014acoustic,sindona2016plasmon}, this anisotropic behavior is outside the Dirac-cone  approximation, and yields markedly different extrinsic plasmon responses at probing energies smaller than $1.0$~eV.

\subsection{TDDFT+RPA approach\label{3Dvs2Dapproaches}}
Plasmons in solid state materials are typically triggered by electron-beam radiation or photo-currents below a few eV, and charged ions with incident kinetic energy of the order of 0.1-1~keV\cite{Riccardi2003339,PhysRevB.72.165419,Sindona20072805,RiccardiJPhys2010,Cupolillo20101029,RiccardiAPL2012,Ligato201440,Pisarra2014796}.
In the present context, we consider introducing an electron or a photon, with incident momentum ${\bf q}$ and frequency $\omega$, which weakly perturbs the KS electrons of silicene.
The unperturbed density-density response function of the system to the test particle is provided by the Adler-Wiser formula~\cite{kubo1957statistical,adler1962quantum,wiser1963dielectric}.
The latter is generally represented in the reciprocal space as follows
\begin{align}
\label{AdlWi}
&\chi _{{\bf G} {\bf G}'}^{0} =
\frac{2}{N \Omega_0}
\sum_{{\bf k},\nu,\nu'}
\frac{
(f_{\nu {\bf k}}-f_{\nu' {\bf k}+{\bf q}})
\rho_{\nu\nu'}^{{\bf k}{\bf q}}({\bf G})\,
\rho_{\nu\nu'}^{{\bf k}{\bf q}}({\bf G}')^{\ast}
}{
\omega+\varepsilon _{\nu{\bf k}}-\varepsilon_{\nu'{\bf k}+{\bf q}}+ i \;\eta }.
\end{align}
Here, the factor of $2$ accounts for the spin degeneracy, $f_{\nu {\bf k}}$ labels the Fermi-Dirac distribution,
\Ignore{
\begin{align}
\rho_{\nu\nu'}^{k{\bf q}}({\bf G})=&\int_{\Omega} d^3r \braket{\nu {\bf k}}{{\bf r}}
e^{- i ({\bf q}+{\bf G})\cdot\mathbf{r}}
\braket{{\bf r}}{\nu'{\bf k}+{\bf q}}\notag \\
=&\sum_{\bf G'} c_{\nu {\bf k}-{\bf G}+{\bf G'}} c_{\nu' {\bf k}+{\bf q}+{\bf G}'},
\label{rhocc}
\end{align}
}
\begin{align}
\rho_{\nu\nu'}^{k{\bf q}}({\bf G})=&\int_{\Omega} d^3r \braket{\nu {\bf k}}{{\bf r}}
e^{- i ({\bf q}+{\bf G})\cdot\mathbf{r}}
\braket{{\bf r}}{\nu'{\bf k}+{\bf q}}
\label{rhocc}
\end{align}
are density-density correlation matrix elements, and
$\eta$ yields a positive broadening~(chosen to the value of $0.02$~eV\Ignore{, which gives well converged results}).

The interacting density-density response function \Ignore{(or the susceptibility) of silicene} stems from the central equation of TDDFT~\cite{petersilka1996excitation,onida2002electronic}\Ignore{, which has the form}
\begin{equation}
\chi_{{\bf G}{\bf G}'}= \chi^0_{{\bf G}{\bf G}'} + (\chi^0 v \chi)_{{\bf G}{\bf G}'},
\label{chi}
\end{equation}
where $v_{{\bf G}{\bf G}'}$ represent the Fourier coefficients of an effective electron-electron interaction.
In the RPA, these terms are approximated to the bare Coulomb potential:
\begin{equation}
v_{{\bf G}{\bf G}'} \approx \Ignore{v^0_{{\bf G}{\bf G}'}=}4\pi\delta_{{\bf G}{\bf G}'}/|{\bf q} + {\bf G}|^2,
\label{v0Coul}
\end{equation}
where a purely 3D periodic system is taken into account.
The drawback of this approach is the non-negligible interaction between the periodic replicas, no matter how large their relative distance is.
To cancel out this unphysical phenomenon, basically due to the long-range character of the Coulomb potential, we replace Eq.~\eqref{v0Coul} by the truncated Fourier integral over the out-of-plane axis ($z$):
\begin{equation}
\label{eq3}
v_{{\bf G}{\bf G}'} \approx \frac{2\pi}{|{\bf q} + {\bf g}|}\int_{-L/2}^{L/2} dz \int_{-L/2}^{L/2} dz' e^{ i  ({G_z} z- {G'_z} z')-|{\bf q} + {\bf g}||z + z'|},
\end{equation}
where ${\bf g}$ and ${G_z}$ denote the in-plane and out-of-plane components of  ${\bf G}$.
Eq.~\eqref{eq3}, tending to Eq.~\eqref{v0Coul} for $L\to\infty$, has been proved to establish a proper two-dimensional cut off~\cite{despoja2012ab,despoja2013two,novko2015changing,pisarra2016dielectric}, eliminating the interaction between charge densities located at the different slabs\Ignore{~(being mostly due by the $v^0_{{\bf 0}{\bf 0}}$ term)}.

Once the interacting matrix elements  in Eq.~\eqref{chi} have been set by Eq.~\eqref{eq3}, \Ignore{by letting $w_{{\bf G}{\bf G}'}=v_{{\bf G}{\bf G}'}$,}  the inverse dielectric matrix is obtained as
\begin{equation}
(\epsilon^{-1})_{{\bf G} {\bf G}'}=\delta_{{\bf G}{\bf G}'}+(v \chi)_{{\bf G} {\bf G}'}.
\label{eq4}
\end{equation}
Collective excitations~(plasmons) are established by the zeros in the real part of the macroscopic dielectric function (permittivity) $\epsilon^M=1/(\epsilon^{-1})_{\mathbf{0}\mathbf{0}}$, whose imaginary part gives the optical absorption spectrum.

The plasmon structure is provided by the energy-loss function, being proportional to the imaginary part of the inverse permittivity:
\begin{equation}
E_{\msc{loss}}=-\im[(\epsilon^{-1})_{\mathbf{0}\mathbf{0}}].
\label{eq6}
\end{equation}
Nonlocal field effects are included in Eq.~\eqref{eq6} through the off-diagonal elements of $\chi_{{\bf G}{\bf G}'}$~\cite{kramberger2008linear}.
As in MG and BLG, we verified that $51$ $\mathbf{G}$ vectors of the form $(0,0,G_z)$, sorted in length order from smallest to largest, lead to well-converged results\Ignore{, in the frequency-momentum region considered in this study}.

In what follows we discuss the dielectric and  energy-loss  properties of silicene, as obtained from the outlined TDDFT+RPA scheme at room temperature and under several intrinsic or extrinsic conditions.

\section{Intrinsic silicene\label{IntrinsicSil}}

The ground-state electron density for the optimized lattice constant $a=3.82$~{\AA} and buckling parameter $\Delta=0.45$~{\AA}~(Sec.~\ref{DFTCalcs}) was used--in a non self consistent run--to improve the resolution on the KS eigensystem $\{\ket{\nu {\bf k}},\varepsilon_{\nu {\bf k}}\}$.
The latter was recalculated on an MP mesh of ${\NMP}=360\times360\times1$ ${\bf k}$ points, including up to $60$~bands.
This result was plugged in the TDDFT+RPA machinery, summarized by Eqs.~\eqref{AdlWi}-\eqref{eq6}, to have an accurate representation of the electronic excitations and energy loss properties up to $20$~eV.
Similar calculations were run for the lattice-constant values $a = 3.86$ and $3.89$~{\AA}, as reported in Appendix~\ref{appI}.

\begin{figure}[b]
\centering{
\includegraphics[width=0.495\textwidth]{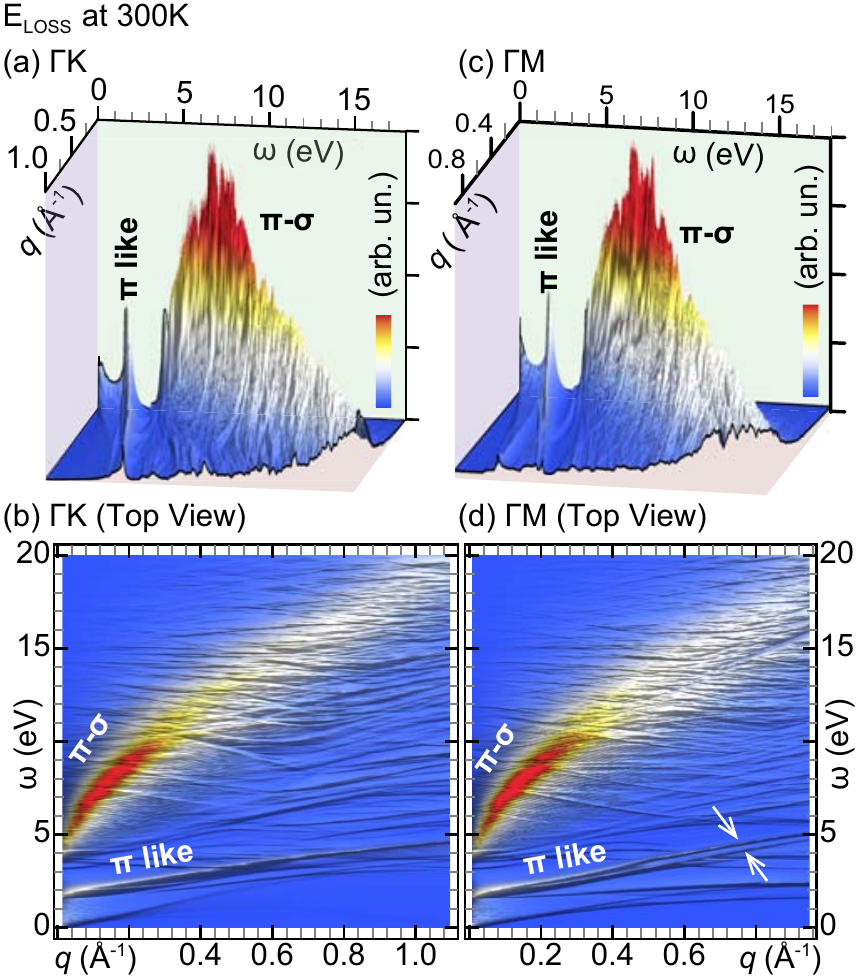}
}
\vskip -14pt
\caption{Energy-loss function of undoped silicene at $T=300$~K vs $\omega{\leq}20$~eV and ${\bf q}$ along the full ${\Gamma}K$~[(a),(b)] and  ${\Gamma}M$~[(c),(d)] paths of Fig.~\ref{bandSi}.
In (b) and (d), the intensity scale is cut at $70\%$ of the peak maximum, due to the $\pi$-$\sigma$ plasmon~(as also shown in Fig.~\ref{3Dloss_line}).
\label{lossHighE}
}
\end{figure}

The  energy-loss  spectrum of undoped~(intrinsic) silicene, with the LDA optimized geometry, is reported in Fig.~\ref{lossHighE} for a broad range of frequencies, ranging from the lower THz to the \Ignore{extreme} UV, and incident momenta along $\Gamma$K and $\Gamma$M.
Two plasmon structures can be clearly distinguished, i.e., a $\pi$~like plasmon and a $\pi$-$\sigma$ plasmon that are peaked  respectively at $\omega \sim 1.7$~eV and $\omega \sim 4$~eV, in correspondence with the lowest sampled $q$ values, being smaller than $10^{-2}$~{\AA}$^{-1}$.
The outlined interband modes are counterparts to the well-known $\pi$ and $\pi$-$\sigma$ plasmons of MG, BLG, fewlayer graphene~ (i.e., five to ten stacked MG sheets) and graphite~\cite{eberlein2008plasmon,despoja2012ab,despoja2013two,pisarra2014acoustic,pisarra2016dielectric,sindona2016plasmon},
sharing with the former a squareroot behavior of the energy-momentum dispersions in the optical region.
Nevertheless, the reduced width of the $\pi$~($\pi^{*}$) and $\sigma$~($\sigma_d^{*}$) bands~[Fig.~\ref{bandSi}(b)] and  the peculiar DOS features~[Fig.~\ref{bandSi}(c)], which characterize silicene with respect to MG, cause a red-shift and a different relative weight of the intrinsic plasmon peaks.

Indeed the $\pi$ and $\pi$-$\sigma$ structures of MG have similar peak intensities~\cite{eberlein2008plasmon}, whereas the $\pi$-$\sigma$ peak of silicene is generally larger (by a factor of 3 at $q > 0.1$~{\AA}$^{-1}$) than the $\pi$ like peak.
This fact can be ascribed to the weakening of the $\pi$ bonds in silicene due to the mixed $sp^2$-$sp^3$ hybridization discussed in Sec.~\ref{DFTCalcs}.
Another distinctive aspect of the intrinsic response of silicene is the double (quasilinear) dispersion of the $\pi$~like plasmon in the energy-momentum region $\omega \gtrsim 2.5$~eV and $q \gtrsim 0.3$~{\AA}$^{-1}$, along $\Gamma$M~[Fig.~\ref{lossHighE}(c) and \ref{lossHighE}(d)].
In this region, the collective oscillation  is associated to SP excitations between high DOS points connecting the $\pi$ and $\sigma_d^*$ or $\pi^*$ bands~[Fig.~\ref{bandSi}(c)].

\begin{figure}[h]
\centering{
\includegraphics[width=0.495\textwidth]{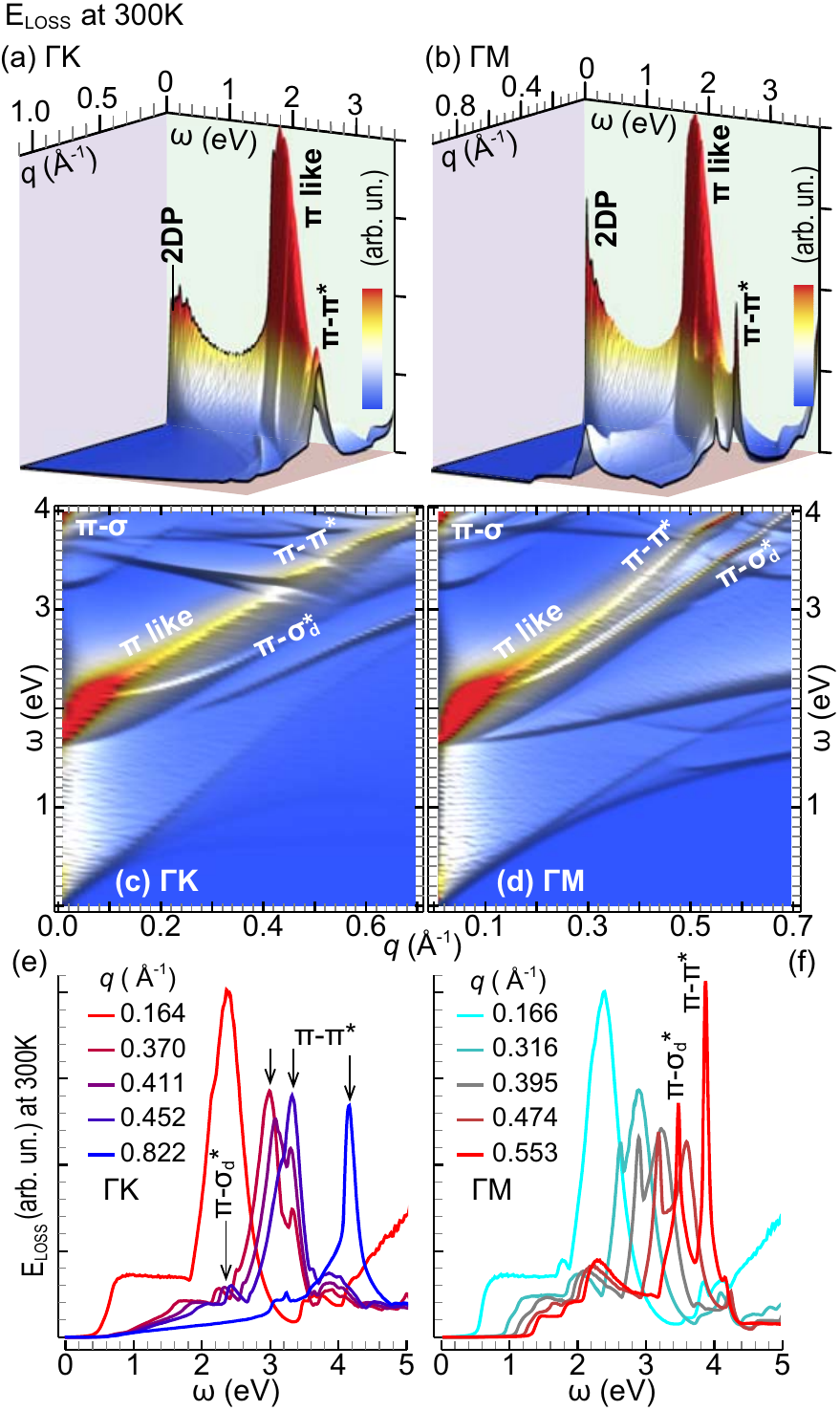}
}
\vskip -14pt
\caption{${E}_{\msc{loss}}$ of undoped silicene vs $\omega<5$~eV and momentum $q<0.7$~\AA$^{-1}$ along  ${\Gamma}K$ [(a),(c),(e)] and ${\Gamma}M$ [(b),(d),(f)).
The intensity scale in (c), (d) is cut at $24\%$ of the peak maximum, occurring at low $q$ and low $\omega$ [shown in (a), (b)]
\label{zoomEL}
}
\end{figure}

A more detailed view on the $\pi$~like plasmon is offered by Fig.~\ref{zoomEL}, where we see how the very close (or overlapping) energy levels in the $\pi^*$ and $\sigma_d^*$ bands~[Fig.~\ref{bandSi}(c)] generate distinct plasmon features at high-DOS points~[Fig.~\ref{bandSi}(d)].
For small incident momenta around the $M$ point~($q<0.1$~{\AA}$^{-1}$) and energies below $2.5$~eV, the large $\pi^*$ DOS peak hides the $\sigma_d^*$ contribution~[Fig.~\ref{bandSi}(d) and~\ref{bandSi}(f)], and a single plasmon character dominates along both $\Gamma$K and $\Gamma$M.
As $q$ increases above $\sim 0.3$~{\AA}$^{-1}$ and $\omega$ gets larger than $3$~eV, the $\sigma_d^*$ component increases becoming of the same order as the $\pi^*$ component~[Fig.~\ref{bandSi}(d) and~\ref{bandSi}(f)].
This increase is associated to a larger splitting between the antibonding bands, which
leads to a well-resolved two-peak structure in the energy-loss spectra.
The latter presents markedly distinct features along $\Gamma$K and $\Gamma$M, being a signature of the deeply anisotropic character of the dielectric response of the system.
In either cases, the $\pi$~like plasmon is indeed a hybridized plasmon where the role of $\pi$-$\pi^*$ and $\pi$-$\sigma^*$ components, i.e., the relative spectral weight of the associated SP processes, is modulated by the excited electronic structure.
In addition, slight changes in the lattice constant cause some distortions of the $\pi$~like plasmon peaks without altering the peak ratio of the $\pi$-$\pi^*$ and $\pi$-$\sigma^*$ parts~(Appendix.~\ref{appI}).

Besides the $\pi$~like and $\pi$-$\sigma$ plasmons, Fig.~\ref{zoomEL} also shows a nonnegligible intraband plasmon, peaked at energies below $1$~eV, which is generated by a conduction electron concentration $n^*$ of $2.9{\times}10^{11}$~cm$^{-2}$ at room temperature.
This value is larger than the one found in  MG, because  of  the  smaller  Fermi-velocity values that characterize silicene, in spite of its larger unit-cell area,  which correspond to a lower slope in the vanishing DOS at the K points (Figs.~\ref{bandSi} and~\ref{FVeloc}).
The intraband feature is barely visible in  Figs.~\ref{lossHighE} and~\ref{zoomEL}, where it is hidden by the onset of the $\pi$ like structure\Ignore{ and the energy resolution~(0.01~eV) used to sample the $0$-$20$~eV range}.
However, it can be detected at the lowest $q$'s in Figs.~\ref{ReImLoss} and Figs.~\ref{3Dloss_line} (of  Appendix~\ref{appI}), where an extra peak below $1$~eV is clearly spotted in both ${E}_{\msc{loss}}$ and $\im(\epsilon^M)$.
\redc{Hence, intraband plasmons are also
possible in the intrinsic limit when finite temperature is considered.}

\begin{figure}[h]
\centering{
\includegraphics[width=0.485\textwidth]{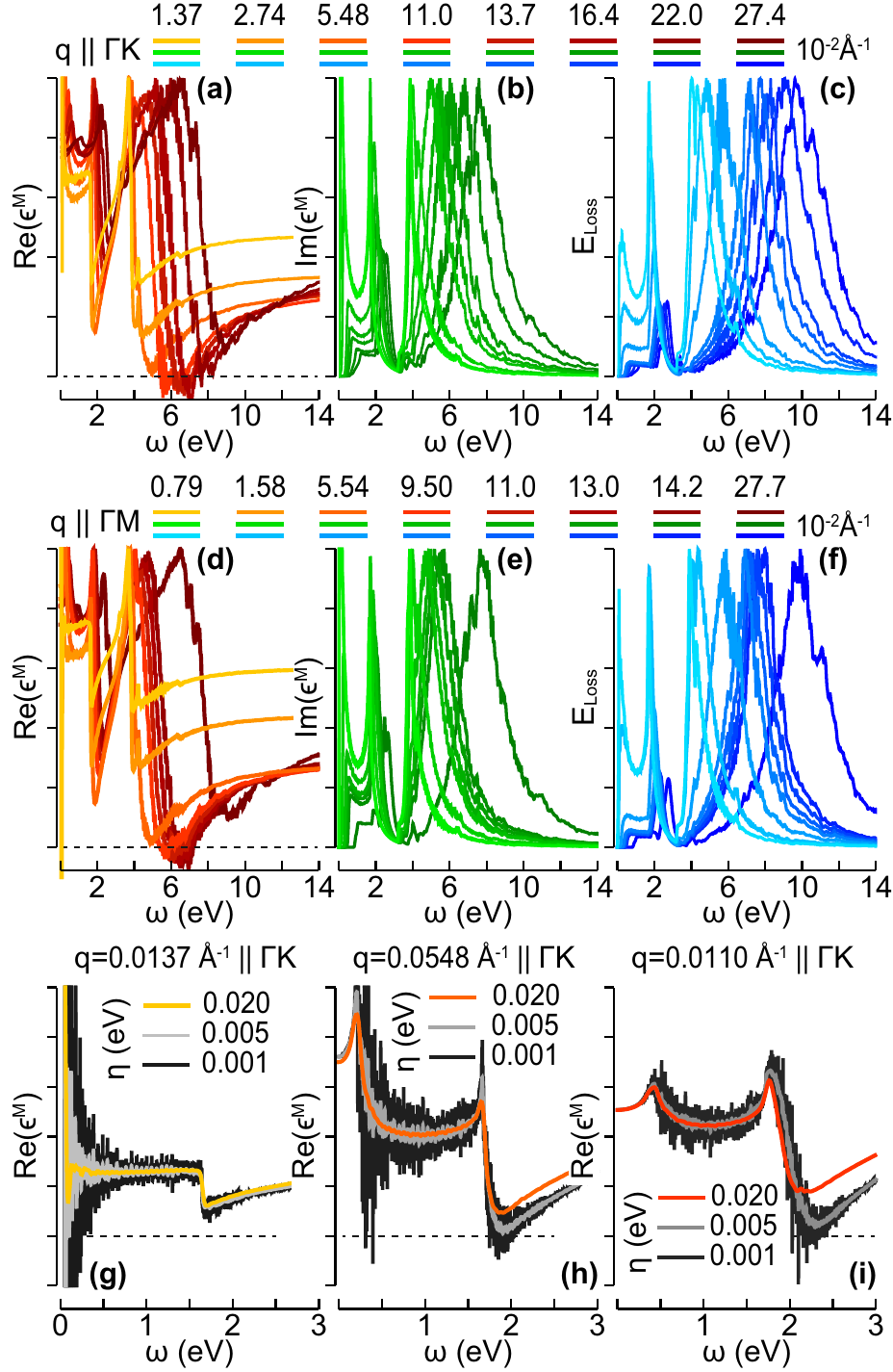}
}
\vskip -14pt
\caption{Complex permittivity and energy-loss function of intrinsic silicene for selected transferred momenta below $0.3$~{\AA}$^{-1}$ along $\Gamma{K}$ and $\Gamma{M}$.
In (a)-(f), $\epsilon^M$ is normalized to the largest peak of its imaginary part, while ${E}_{\msc{loss}}$ is normalized to its maximum value, with the infinitesimal broadening width $\eta$ of Eq.~\eqref{AdlWi} being fixed to $0.02$~eV.
In (g)-(i), $Re(\epsilon^M)$, normalized to its zero frequency peak, is reported for three different values of $\eta$, to show how the $Re(\epsilon^M)$ tends to have a zero, in coincidence with the intraband and $\pi$~like plasmon resonances, as $\eta$ artificially gets smaller and smaller.
\label{ReImLoss}
}
\end{figure}
From a closer analysis of $\epsilon^M$, we realize that only the $\pi$-$\sigma$ peak is a well-defined plasmon in a specific region of the energy-momentum domain, with $q \sim 0.1-0.2$~{\AA}$^{-1}$ and $\omega\sim 5-7$ eV, where it corresponds to a zero in  $\re({\epsilon}^M)$ at a frequency where $\im({\epsilon}^M)$ is small.
All other $\pi$~like and $\pi$-$\sigma$ structures can still be treated as collective excitations, superimposed to SP processes between high-DOS points, whose average lifetime broadening~($\eta=0.02$~eV in Eq.~\eqref{AdlWi}) induces a large Landau damping.

A parallel situation has been observed in MG~\cite{despoja2013two,pisarra2014acoustic,novko2015changing,pisarra2016dielectric}.
Indeed, as we let the lifetime broadening parameter tend artificially to zero, the condition for occurrence of a collective oscillation becomes more and more likely to be satisfied.
In support of this argument, Figs.~\ref{ReImLoss}(g)-\ref{ReImLoss}(i) display $\re({\epsilon}^M)$ for $\eta=0.02$, $0.005$ and $\eta=0.001$~eV.
The intraband and $\pi$-like plasmon structures are evidently related to quasi-zeros in $\re({\epsilon}^M)$.
Interestingly enough, the complex permittivity and  energy-loss  function display a nearly isotropic behavior for small values of the transferred momentum, below $0.15$~{\AA}$^{-1}$, while as $q$ gets larger than  $0.15$~{\AA}$^{-1}$ the dielectric response of the system acquires a tensor character.
Similar trends have been reported for MG \cite{kramberger2008linear,pisarra2014acoustic} and BLG~\cite{pisarra2016dielectric}.
As for the absorption spectrum, $\im({\epsilon}^M)$  reflects the main features of the energy-loss function with two or three optical peaks (including the intraband mode at low $q$) being clearly spotted.
The interband peak positions at energies larger than $1$~eV are in agreement with previous absorbance calculations~\cite{matthes2013universal,matthes2014optical}\Ignore{,which demonstrated that $\pi$~like structure is due to strong interband transition around the $M$ point, and
the $\pi$-$\sigma$ feature corresponds to interband transitions along the ${\Gamma}M$ path}.
We should point out that our predictions are based on the two-dimensional cut off procedure outlined in the previous section, while the use of the bare Coulomb interaction would lead to an erroneous positioning of the optical plasmon-peaks~(see Appendix~\ref{appII}).

\section{Extrinsic silicene\label{ExtrinsicSil}}

We now move to \Ignore{our primary focus,} the loss properties of extrinsic silicene with LDA-optimized geometry. In Appendix~\ref{appI}, we will discuss how the effect of stretching the lattice constant (from $a=3.82$ to $a = 3.86$ and $3.89$~{\AA}) does not alter the extrinsic plasmon features.
In what follows, we specifically consider four different charge-carrier concentrations, inducing negative and positive Fermi energy shifts $\Delta E_F$, in the range of $-0.4$ to $0.4$~eV, relative to the Dirac-point energy.
To achieve these extrinsic conditions, we adjusted the occupation factors $f_{\nu {\bf k}}$ and $f_{\nu' {\bf k}+{\bf q}}$ in Eq.~(\ref{AdlWi}) by shifting the Fermi-energy values by $\Delta E_F=\pm 0.2$ and $\pm 0.4$~eV, respectively.
A summary of the sampled doping levels--and corresponding conduction-electron or valence-hole densities--is given in  Table~\ref{tabdop}.
The $\pi$~like and $\pi$-$\sigma$ plasmons of the previous section were found to be rather insensitive to the simulated extrinsic conditions, as already indicated by several studies on graphene-related systems~\cite{pisarra2014acoustic,pisarra2016dielectric,vacacela2016gnr,vacacelaBJNANO2017}.
Then, we present a zoomed view of the energy-momentum region $\omega \leq 2.5$~eV, $q < 0.35$~{\AA}$^{-1}$, which was computed with the TDDFT+RPA method illustrated in Sec.~\ref{3Dvs2Dapproaches}, using an equilibrium electronic structure represented on an MP grid of ${\NMP}=720\times720\times1$ ${\bf k}$ points and $12$ bands.

\begin{table}[h]
\centering
\caption{
Fermi-level shifts $\Delta E_F$ induced by adding~(+) or removing(-) $\bar{n}_0$ electrons per unit cell,
with positive or negative charge-carrier concentrations $n_0^*$, which correspond to charge-carrier concentrations $n^*$ at $T=300$ K.}
\begin{tabular}{c c c c}
\hline\hline
$\Delta E_F$ (eV) & $\bar{n}_0$ (el per uc)& $n_0^{*} \;(10^{13}\times\mathrm{cm}^{-2})$  & $n^{*} \;(10^{13}\times\mathrm{cm}^{-2})$\\
\hline
     -0.4 &-0.0510          & -4.039     & -4.098 \\
     -0.2 &-0.0126          & -1.000     & -1.055\\
\Pm 0.0 & -                  & -             &\Pm 0.029 \\
\Pm 0.2 &\Pm 0.0135    &\Pm 1.071 &\Pm  1.136 \\
\Pm 0.4 & \Pm 0.0597   &\Pm 4.727 &\Pm 4.863 \\
\hline\hline
\end{tabular}
\label{tabdop}
\end{table}

\begin{figure}[h]
\centering{
\includegraphics[width=0.48\textwidth]{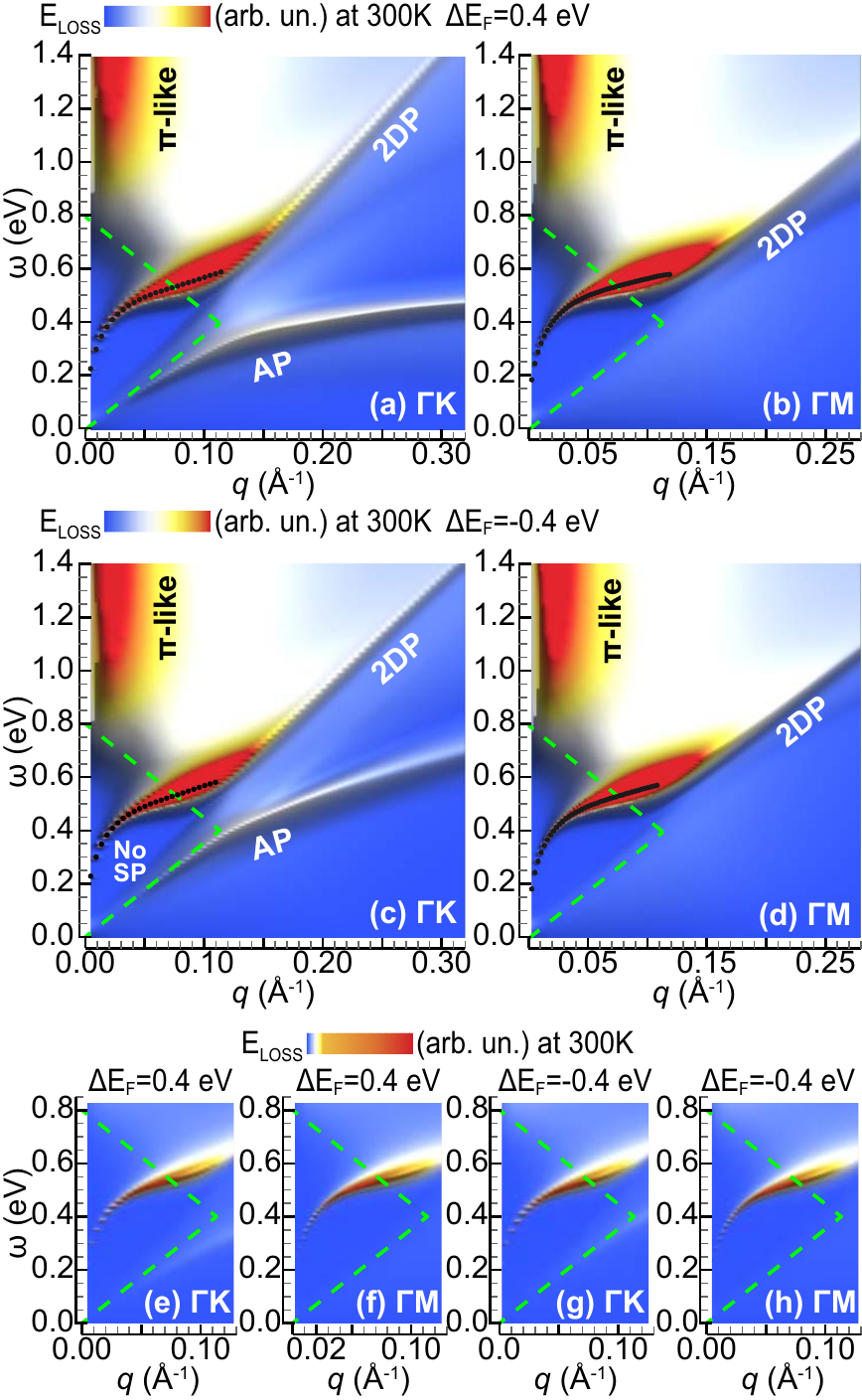}
}
\vskip -14pt
\caption{Energy-loss function at room temperature of positively or negatively doped silicene for $\Delta E_F=\pm0.4$ eV, i.e.,
${E}_{\msc{loss}}$  vs $\omega<1.2$~eV and $q<0.27$~{\AA}$^{-1}$ along
${\Gamma}K$ [(a), (c)] and  ${\Gamma}M$ [(b), (d)].
The intensity scale is cut at $5\%$ of the $\pi$~like plasmon peak, visible at $q<{0.05}~$\AA$^{-1}$ and $\omega > 0.8$~eV, which corresponds to $0.1\%$ of the 2DP peak.
The black dots represent the $(q,\omega)$ values at which $\re({\epsilon}^M)=0$ and $\im({\epsilon}^M)$ is zero or small.
These mostly fall in a region where SP excitations are absent, where the 2DP looks like a non-Landau damped  peak.
An insight onto the no-SP-excitation region (delimited by dashed green lines) is given in (e)-(h), where the intensity scale is cut at $3\%$ of the 2DP peak.
\label{Dop04}
}
\end{figure}

\begin{figure}[h]
\centering{
\includegraphics[width=0.48\textwidth]{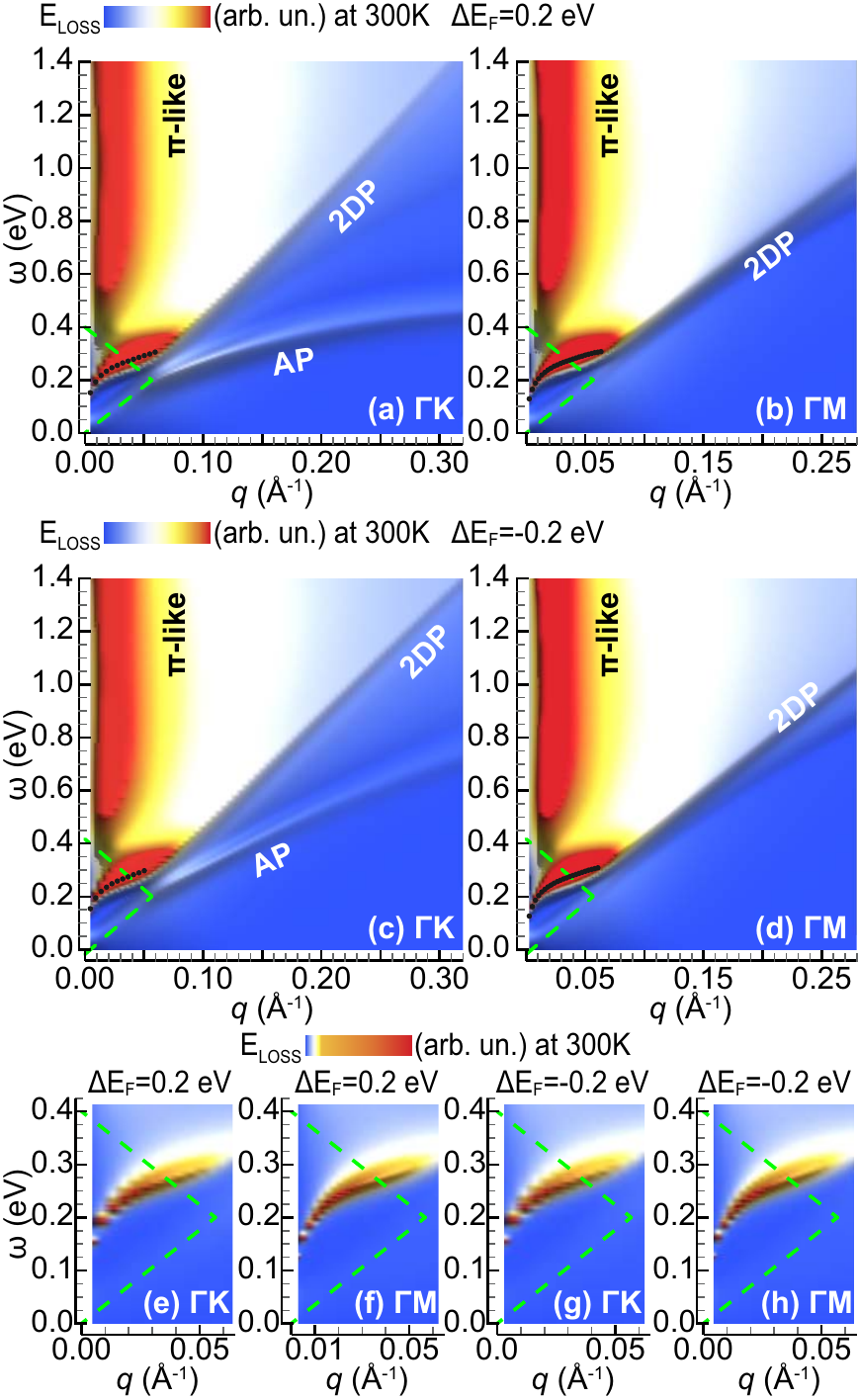}
}
\vskip -14pt
\caption{Energy-loss function at room temperature of positively or negatively doped silicene, for $\Delta E_F=\pm0.2$ eV, along ${\Gamma}K$ [(a), (c), (e), (g)] and  ${\Gamma}M$ [(b), (d), (e), (h)].
As in Fig.~\ref{Dop04}, the intensity scale is cut at $5\%$ of the $\pi$~like plasmon peak in (a)-(d), and at  $3\%$ of the 2DP peak in (e)-(h).
The black dots denote the zeros of $\re({\epsilon}^M)$ in a region where $\im({\epsilon}^M)$ is zero or small.
The dashed green lines delimit the no-SP-excitation region, where the 2DP is not affected by Landau damping.
\label{Dop02}
}
\end{figure}

The energy-loss  spectra for $\Delta E_F=\pm0.4$~eV and $\Delta E_F=\pm0.2$~eV are shown as density plots in Figs.~\ref{Dop04} and~\ref{Dop02}, respectively.
The most striking feature here is the appearance of two distinct plasmon resonances, which are nearly absent in intrinsic silicene at room temperature~[Fig.~\ref{zoomEL}(a) and~\ref{zoomEL}(c)].

The most intense peak is associated to a two-dimensional plasmon, denoted 2DP, which is clearly manifested along both $\Gamma$K and $\Gamma$M, and exhibits a $q^{1/2}$ dispersion at optical wave lengths, as the conventional plasmon of a two-dimensional electron gas.
This mode has been predicted and analyzed in a number of theoretical studies on extrinsic MG, ranging from two-band models in the Dirac cone approximation~\cite{Sarma.PhysRevB.75.205418} to TDDFT approaches~\cite{pisarra2014acoustic,sindona2016plasmon,despoja2013two}, and it is at the heart of technological applications in  graphene plasmonics~\cite{ju2011graphene,bao2012graphene,garcia2014graphene}.

Despite the difference in intensity, a second plasmon of acoustic nature, denoted AP, is clearly visible for momentum transfers along $\Gamma$K, being generated by the two type of Dirac electrons responsible for the different Fermi-velocity values along $\Gamma{K}$~[Fig.~\ref{FVeloc}(a) and~\ref{FVeloc}(b)].
The 2DP corresponds to the two types of electrons oscillating in-phase with one another, and the AP mode corresponds to electrons oscillating out-of-phase.
This mode has been originally identified in MG~\cite{pisarra2014acoustic,sindona2016plasmon} and offers a linear energy-dispersion in the low-$q$ limit.
\begin{figure}[h]
\vskip -10pt
\centering{
\includegraphics[width=0.48\textwidth]{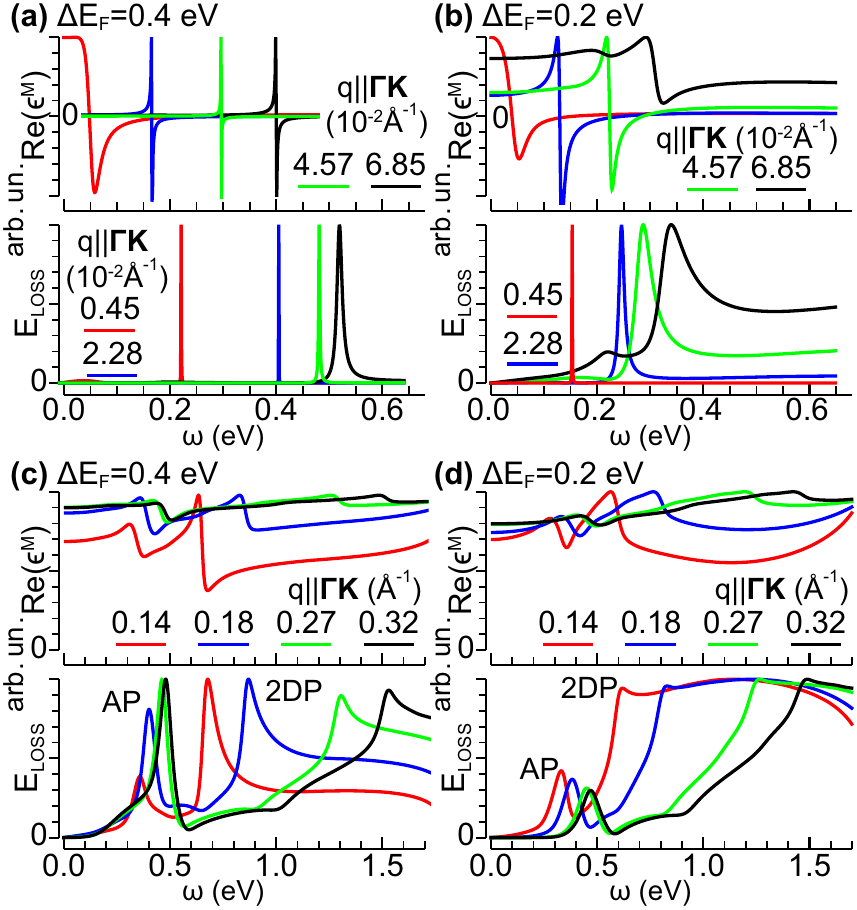}
}
\vskip -16pt
\caption{$\re({\epsilon}^M)$ and ${E}_{\msc{loss}}$--normalized to their maximum values--for extrinsic silicene at $\Delta E_F=0.4$~eV [(a),(c)] and $\Delta E_F=0.2$~eV [(b),(d)] for some fixed ${\bf q}$ points along $\Gamma{K}$.
\label{exDEF2}
}
\end{figure}
\begin{figure}[h]
\centering{
\includegraphics[width=0.48\textwidth]{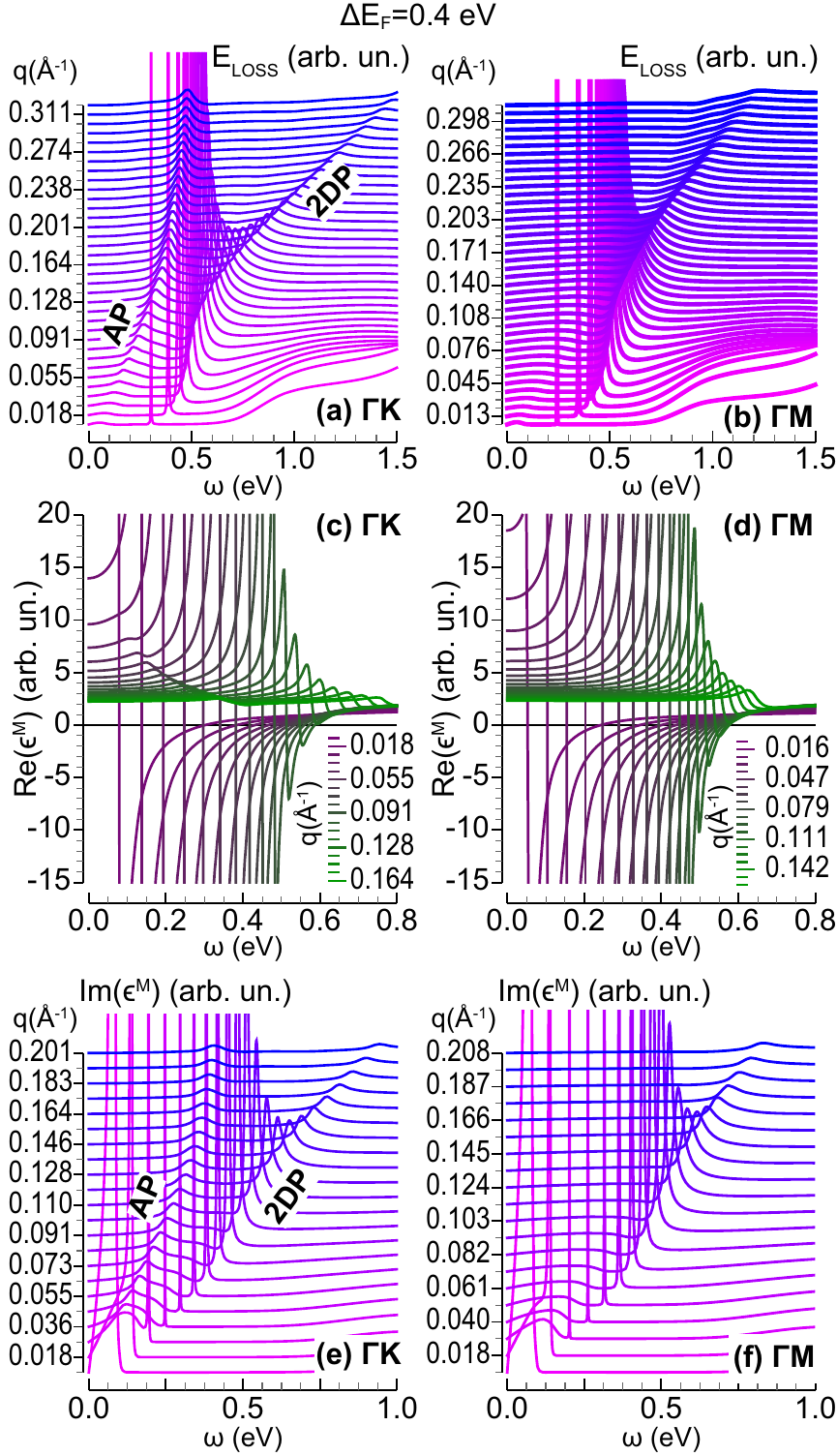}
}
\vskip -14pt
\caption{Dielectric properties of extrinsic silicene for $\Delta E_F=0.4$ eV, i. e., $\re({\epsilon}^M)$, $\im({\epsilon}^M)$ and ${E}_{\msc{loss}}$ vs $\omega\leq1$~eV, for $q \lesssim 0.3$~\AA$^{-1}$ along ${\Gamma}K$ [(a), (c), (e)] and ${\Gamma}M$ [(b), (d), (f)].
\label{exDEF}
}
\end{figure}

A signature of the asymmetric dielectric response of the material is provided by the observation that the 2DP mode is present along both $\Gamma$K and $\Gamma$M, whereas the AP mode is present only along $\Gamma$K.
This means that the widely used Dirac cone approximation is capable of reproducing only the 2DP features.
By comparing Figs.~\ref{Dop04} and~\ref{Dop02}, we see that interplay between the 2DP and the AP is strongly sensitive to the extrinsic conditions, including the doping sign, being associated to different charge-carrier densities~(Table~I).
More importantly, there is a region in the energy-momentum domain, corresponding to $q \sim 0.1-0.2$~{\AA}$^{-1}$, $\omega \sim 0.2-0.6$~eV, where the AP mode is competitive with the 2DP.
Although many recent calculations have reported the existence of two extrinsic plasmons in graphene-related materials~\cite{pisarra2016dielectric,Low.PhysRevLett.112.116801}, the peculiarity of the AP mode in silicene and MG~\cite{pisarra2014acoustic,sindona2016plasmon} is that it occurs in a single, virtually gapless band.
Our study suggests that the occurrence of such a mode is another universal characteristic of the honeycomb lattice~(in addition to the above mentioned infrared absorbance).
It is worth mentioning \Ignore{the fact} that the 2DP and the $\pi$~like structure tend to get closer to each other, as the doping decreases. This is why at room temperature and zero doping the onset of $\pi$~like mode mostly hides the 2DP contribution~[as already pointed out in Sec.~\ref{IntrinsicSil}, with reference to Figs.~\ref{zoomEL} and \ref{ReImLoss}].

Also interesting to notice is the opening of a gap in the SP excitation spectrum for low $q$ and $\omega$, along both $\Gamma$K and $\Gamma$M.
Such a gap is in close agreement with the triangular region  predicted by two-band models in the Dirac cone approximation~\cite{Sarma.PhysRevB.75.205418}.
The latter is delimited the by dashed green lines in Figs.~\ref{Dop04} and~\ref{Dop02}, which correspond to $\omega=v_F q$ and $\omega = 2 \Delta E_F -v_F q$ (with $v_F$ being the average Fermi-velocity introduced in Sec.~\ref{DFTCalcs}).

The 2DP falls onto the no-SP excitation region at the lowest $q$'s, where it appears as a sharp peak not affected by  Landau damping, as displayed in Figs.~\ref{Dop04}(e)-\ref{Dop04}(h), \ref{Dop02}(e)-\ref{Dop02}(h), \ref{exDEF2}(a) and~\ref{exDEF2}(b).
Accordingly, the real permittivity has well defined zeros in this region [reported as black dots in Figs.~\ref{Dop04}(a)-\ref{Dop04}(d) and \ref{Dop02}(a)-\ref{Dop02}(d)].
For $(\omega,q)$ points outside the no-SP excitation region, the 2DP gets more and more damped, with the $\re({\epsilon}^{M})=0$ condition being violated at large $q$ values, as shown in Fig.~\ref{exDEF2}(c) and ~\ref{exDEF2}(d), where the competition mechanism between the 2DP and AP modes is emphasized.
This explains why the intrinsic response of Silicene does not offer well-defined plasmon excitations, with the area of the no-SP-excitation region being virtually zero.
As a further evidence in support of this argument, in Fig.~\ref{exDEF} we show a complete analysis of the macroscopic dielectric function and the two-plasmon structure for $\Delta E_F=0.4$~eV.
Indeed, \Ignore{Figs.~\ref{Dop04}(a), \ref{Dop04}(b), Figs.~\ref{exDEF2}(a), \ref{exDEF2}(b) and} Figs.~\ref{exDEF}(c), \ref{exDEF}(d) prove that the 2DP is a genuine collective oscillation in the no-SP-excitation region.
On the other hand the AP mode lies outside this region, and the corresponding plasmon oscillation is always damped in the Landau sense.

\section{Conclusions}

We have presented a full TDDFT+RPA analysis of the dielectric properties and plasmon dispersion of intrinsic and extrinsic silicene in freestanding form, in absence of external fields and spin-orbit coupling\redc{, suitable for energies above $0.1$~eV and incident momenta larger than $10^{-4}$~\AA$^{-1}$.}
The energy-loss spectra of intrinsic silicene have singled out two interband plasmon structures, lying at energies above 1.5~eV, which resembles the $\pi$ and $\pi$-$\sigma$ modes of MG.
The low-$q$ features of the plasmon peaks have been found in agreement with previous absorbance calculations~\cite{bechstedt2012infrared}.
A more careful analysis has revealed that the $\pi$~like plasmon of silicene is assisted by SP processes between hybridized sp$^2$ and sp$^3$ states, connecting the $\pi$ band to the $\pi^{*}$  and $\sigma^{*}$~like bands, which generates a double energy-momentum dispersion, best resolved for momentum transfers along $\Gamma{M}$.
Similar characteristics are \Ignore{consequence of the hybridized states of silicene, detectable in its ground-state band structure, and are} expected to occur also in other buckled two-dimensional materials in honeycomb geometry, such as germanene.

At lower energies (below $\sim 1$~eV), the energy-loss spectra of extrinsic silicene show two distinct oscillations, whose relative strength can be modulated by the doping concentration of injected or ejected electrons.
Similarly to MG, silicene presents a two-dimensional plasmon, which shares many common features with the plasmon mode of a two-dimensional electron gas, plus an acoustic  mode, being observable only along specific directions of the incident momentum.
Unlike the spin-polarized and valley modes found at meV energies and far-IR momentum transfers~\cite{PhysRevB.89.195410,PhysRevB.89.201411,PhysRevB.90.035142}, being related to the opening of a band gap, these two modes are generated in a gapless band-structure by two different types of charge carriers, i.e., Dirac electrons moving with distinct Fermi velocities.

Our findings suggest that the 2DP and AP modes exist in other two-dimensional materials in honeycomb lattice, such as germanene, making this features independent on the chemistry of the group-IV element, buckling parameter, or hybridization state.
More importantly, they support the argument that silicene-based nanomaterials are excellent options for the design of next-generation nanodevices, in competition with graphene-based nanomaterials.


\noindent\textbf{Acknowledgements} C.V.G. acknowledges the financial support of ``{\em Secretaria Nacional de Educaci\'on Superior, Ciencia, Tecnolog\'ia e Innovaci\'on}'' (SENESCYT-ECUADOR). A. S. acknowledges the computing facilities provided by the  \href{http://www.cineca.it/}{CINECA Consortium}~\cite{GALILEO-SCAI},  within the INF16\_npqcd project under the
  \href{http://www.hpc.cineca.it/news/framework-collaboration-agreement-signed-between-cineca-and-infn}{CINECA-INFN}
 agreement.
\begin{appendix}

\section{Influence of the unit-cell extension on the plasmon structure\label{appI}}
In this first Appendix, we report on how the loss properties of silicene are affected by geometry, specifically focusing on the intrinsic $\pi$~like and extrinsic plasmon features, analyzed in Sec.~\ref{IntrinsicSil} and~\ref{ExtrinsicSil}, respectively.
To this end, we applied the TDDFT+RPA scheme on three different Si-Si bond lengths~(yielding the lattice constant values $a=$3.82, 3.86, 3.89 {\AA} that have been quoted in the literature~\cite{BalendhranSmall2014}.
The buckling parameter has been fixed to its LDA-optimized value in all cases~($\Delta=$0.45~{\AA}).

\begin{figure}[h]
\vskip -10pt
\centering{
\includegraphics[width=0.45\textwidth]{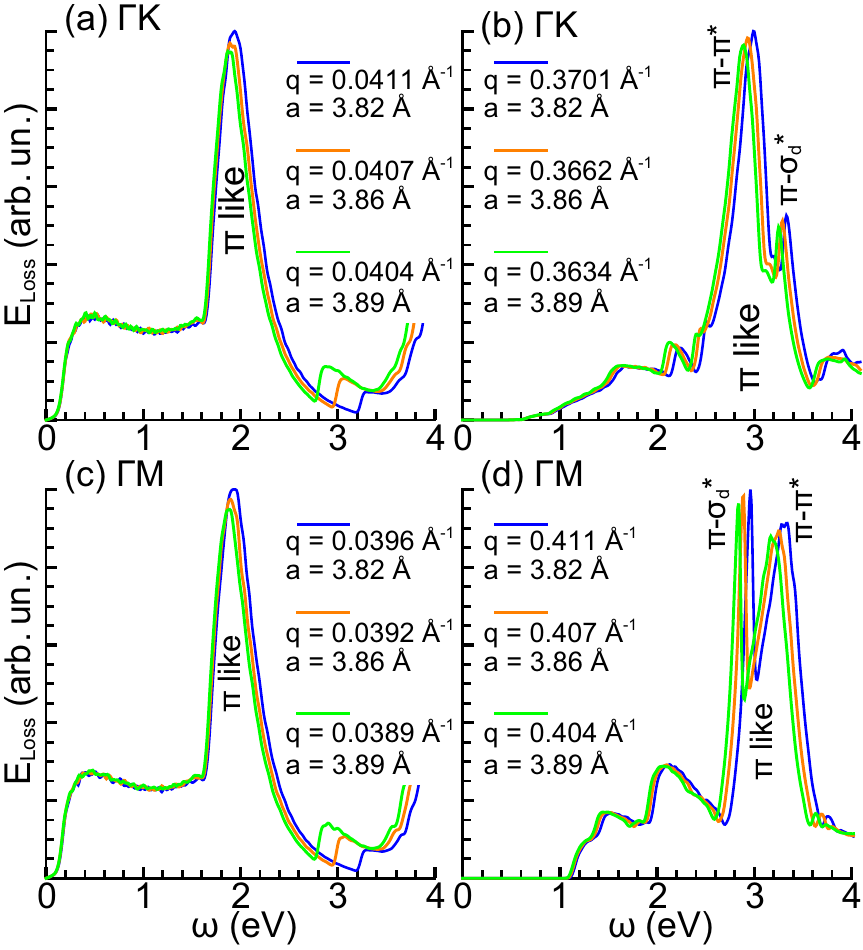}
}
\vskip -16pt
\caption{Energy-loss  function of doped silicene vs $\omega$~$\leq 3$ eV for specific small and large $q$ values (in \AA$^{-1}$) along  ${\Gamma}K$ (a,b) and ${\Gamma}M$ (c,d). Slightly different lattice constant values ($a$=3.82, 3.86, 3.89 \AA) have been tested.
\label{Diff_lattice}
}
\end{figure}
\begin{figure}[h]
\vskip -10pt
\centering{
\includegraphics[width=0.45\textwidth]{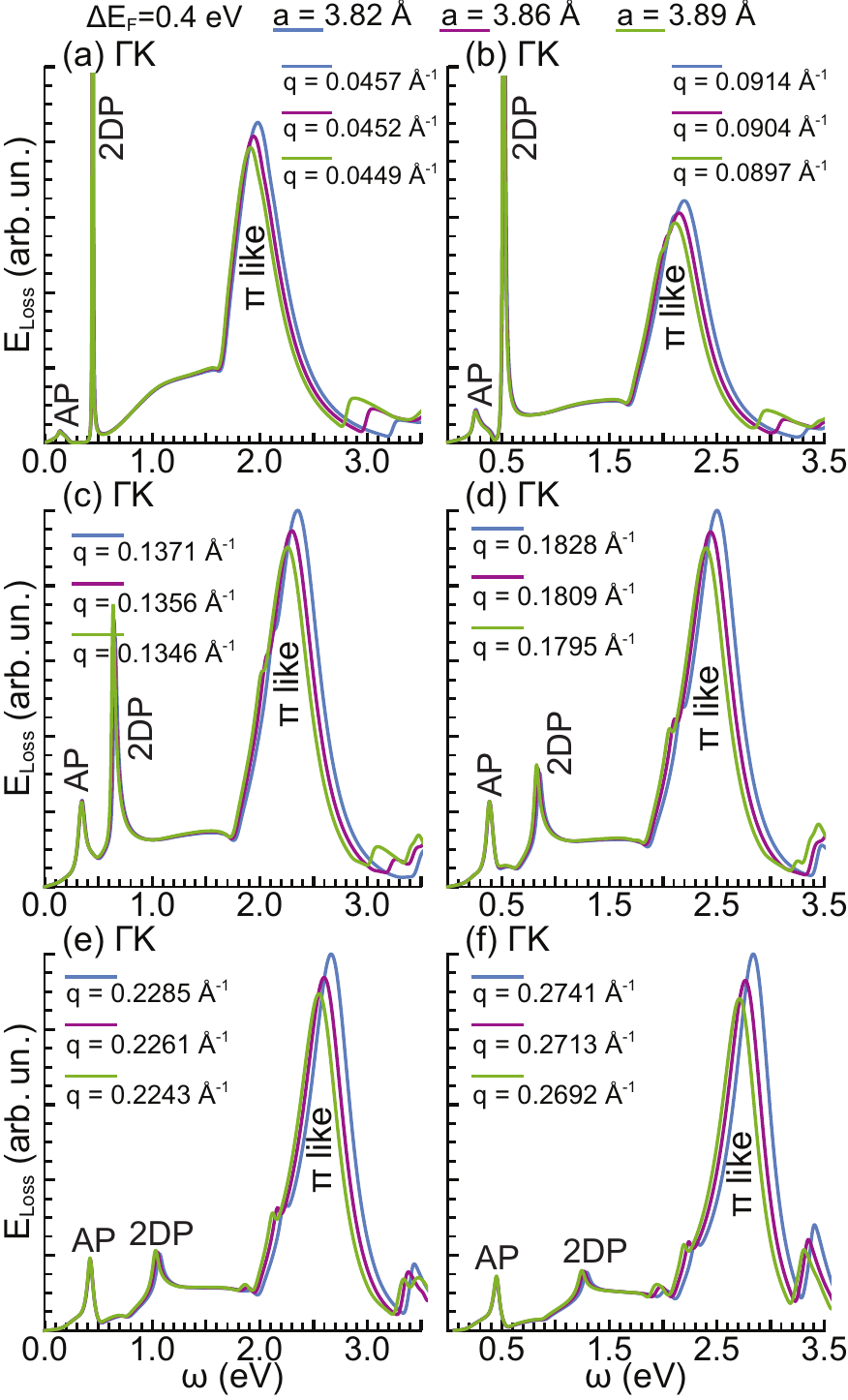}
}
\vskip -16pt
\caption{Energy-loss  function of undoped silicene vs $\omega$~$\leq 3$ eV for specific small and large $q$ values (in \AA$^{-1}$) along  ${\Gamma}K$. Slightly different lattice constant values ($a$=3.82, 3.86, 3.89 \AA) have been tested.
\label{Diff_lattice2}
}
\end{figure}
The calculated loss function is shown in Fig.~\ref{Diff_lattice} for intrinsic silicene and Fig.~\ref{Diff_lattice2} for extrinsic silicene.
The only sensible effect is a red-shift of the $\pi$~like plasmon peaks with the increase of $a$.
In particular, at long wavelengths, peak variations of 3.8$\%$ (along ${\Gamma}K$ for $q = 0.04$~{\AA}$^{-1}$) and 1.3$\%$ (along ${\Gamma}M$ for $q = 0.02$~{\AA}$^{-1}$) are recorded in response to a change in lattice constant of 1.8$\%$.
The same change yields a peak variation of 4$\%$ at small wavelengths along both ${\Gamma}K$ and ${\Gamma}M$ for $q = 0.4$-$0.45$~{\AA}$^{-1}$.
The extrinsic plasmon structure appears to be independent on lattice-constant variations considered here.

\section{Role of the two-dimensional cut off on the Coulomb interaction\label{appII}}
To complete our study, we focus on the long wavelength limit  ($q<10^{−2}${\AA}$^{−1}$)  and  mid infrared  to  near  ultraviolet frequency range ($\omega< 5$ eV) of the plasmon spectra of undoped silicene, and consider some previous theoretical results in comparison with the findings presented in the main text.
\begin{figure}[h]
\vskip -10pt
\centering{
\includegraphics[width=0.48\textwidth]{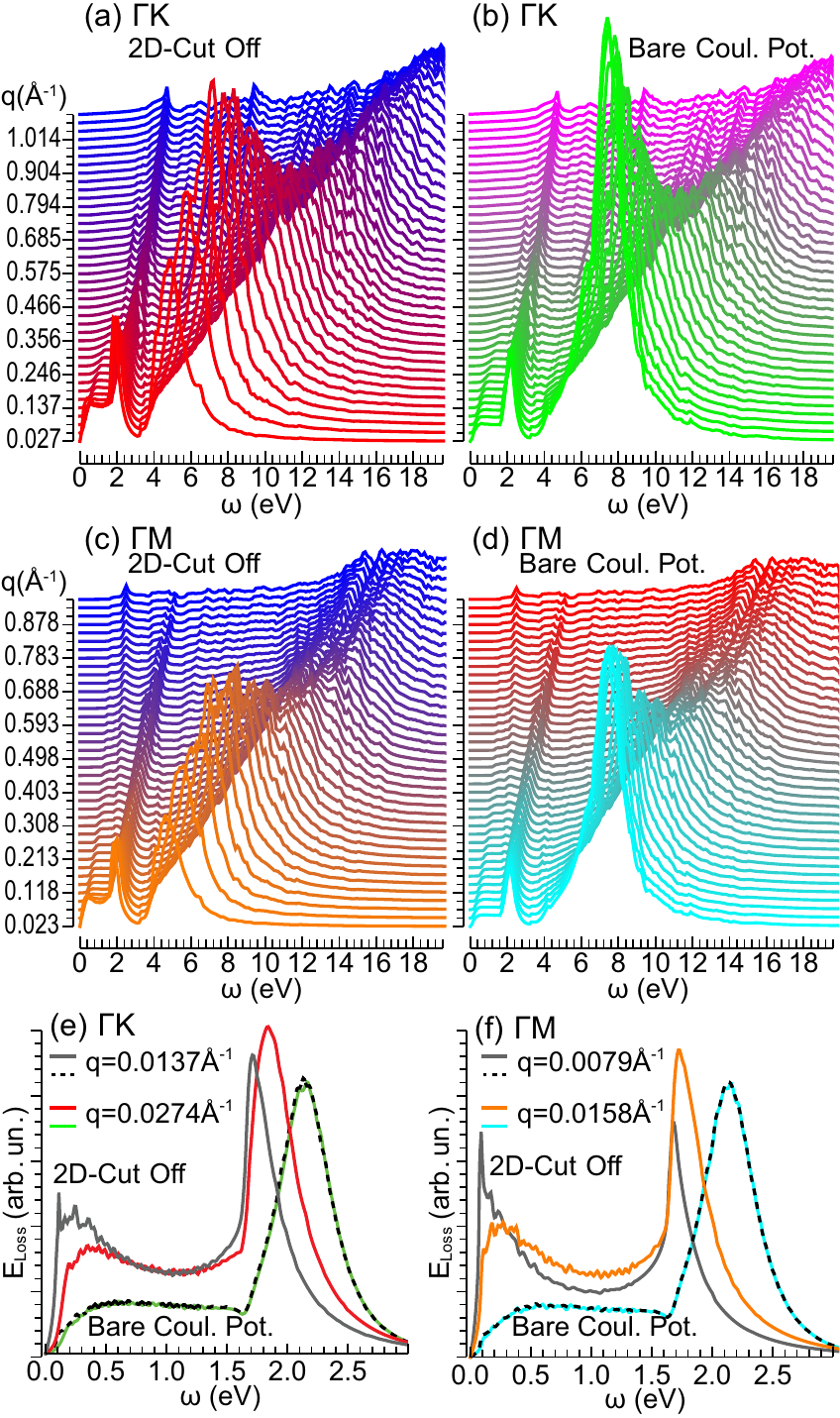}
}
\vskip -16pt
\caption{Energy-loss  function obtained with the TDDFT+RPA method illustrated in Sec.~\ref{3Dvs2Dapproaches}, using the cut off potential of Eq.~\eqref{eq6} [(a),(c)] and the bare Coulomb potential of Eq.~\eqref{v0Coul} [(b),(d)]. The loss curves~(shifted vertically for clarity) are plotted vs $\omega<20$~eV for sampled ${\bf q}$ values (in \AA$^{-1}$) along ${\Gamma}K$ [(a),(b)] and ${\Gamma}M$ [(c),(d)].
The different behavior of the two methods is better illustrated in (e) and (f), where the energy-loss curves are plotted for the lowest two ${\bf q}$ values along ${\Gamma}K$ (e) and ${\Gamma}M$ (f).
\label{3Dloss_line}
}
\end{figure}

Mohan and coworkers~\cite{mohan2013first} have reported two-plasmon features, with  the $\pi$-plasmon lying at 2.16~eV and the $\pi$-$\sigma$-plasmon being peaked at 7.6 eV.
The same peaks have been predicted to be around 2~eV and 9~eV, respectively, by Das {\it et al}~\cite{das2015optical}.
To verify these results, we have solved the Dysonlike equation for the full susceptibility~(Eq.~\eqref{chi}) with the bare Coulomb potential coefficients~(Eq.~\eqref{v0Coul}).
The resulting loss function presents a narrow $\pi$-plasmon peak at 2~eV and a broad $\pi$-$\sigma$-plasmon peak at 7~eV along both $\Gamma$K and $\Gamma$M, at the lowest sampled $q$ values~[Fig.~\ref{3Dloss_line}(b) and~\ref{3Dloss_line}(d)].
On the other hand, with the two-dimensional cut off procedure outlined in the main text the very same two plasmons appear red-shifted~[Fig.~\ref{3Dloss_line}(a) and~\ref{3Dloss_line}(c)], in agreement with the calculations by  Matthes and coworkers~\cite{matthes2013universal,matthes2014optical}.
The disagreement between the two approaches is particularly strong at the lowest sampled ${\bf q}$ points, as attested by the plots of Fig.~\ref{3Dloss_line}(e) and~\ref{3Dloss_line}(f).
This result adds up to a number of previous studies~\cite{despoja2012ab,despoja2013two,novko2015changing,pisarra2016dielectric} where it has been  pointed out that the correct way to investigate and describe the dielectric properties of a two-dimensional system is provided by Eqs.\eqref{AdlWi}-\eqref{eq6}.
The calculations of Fig.~\ref{3Dloss_line} also demonstrate that applying a two-dimensional cut off on the Coulomb potential becomes unnecessary for values of the momentum transfer larger than $0.2$~{\AA}$^{-1}$, which is not surprising because the larger $q$-components of the Coulomb potential do not contribute to coupling of the repeated slab.
\end{appendix}

\clearpage
\end{document}